\documentclass[epjST]{svjour}
\usepackage{graphicx}
\usepackage{url}
\usepackage{hyphenat}
\usepackage{amsmath}
\usepackage{amssymb}
\usepackage{cite}
\usepackage{bm}
\usepackage{color}
\newcommand{\NS}{R}

\begin{document}
\bibliographystyle{epjc}
\setlength{\marginparwidth}{4.2cm}
\title{Optimized GPU simulation of continuous-spin glass models}

\author{Taras Yavors'kii\inst{1} \and Martin Weigel\inst{1,2}}

\institute{Institut f\"ur Physik, Johannes Gutenberg-Universit\"at Mainz, Staudinger
  Weg 7, D-55099 Mainz, Germany \and Applied Mathematics Research Centre, Coventry
  University, Coventry, CV1~5FB, United Kingdom}

\abstract{
  We develop a highly optimized code for simulating the Edwards\hyp{}Anderson
  Heisenberg model on graphics processing units (GPUs). Using a number of
  computational tricks such as tiling, data compression and appropriate memory
  layouts, the simulation code combining over-relaxation, heat bath and parallel
  tempering moves achieves a peak performance of $0.29$ ns per spin update on
  realistic system sizes, corresponding to a more than $150$ fold speed-up over a
  serial CPU reference implementation. The optimized implementation is used to study
  the spin-glass transition in a random external magnetic field to probe the
  existence of a de Almeida-Thouless line in the model, for which we give benchmark
  results.
}

\maketitle


\section{Introduction\label{sec:introduction}}

In spite of a rather substantial concerted research effort extending over more than
three decades, a comprehensive understanding of the theory of the spin-glass state is
still not within reach. Nevertheless, substantial progress has been made since the
suggestion of the Edwards-Anderson Hamiltonian as the standard simplified model of a
spin glass in the 1970s \cite{edwards:75a} and the subsequent formulation of its
highly unusual mean-field theory based on the concept of replica-symmetry breaking
(RSB) \cite{mezard:book}. This concerns, in particular, the nature of the spin-glass
phase itself, for which a number of alternative descriptions have been
proposed. These include a putative continuation of the RSB theory known to be exact
in the mean-field limit to finite-dimensional systems \cite{marinari:00a} as well as
a more conventional picture describing the physics at low temperatures in terms of
droplet excitations similar to those found in ferromagnets
\cite{fisher:86,bray:87a}. In reaction to the conflicting evidence from numerical
calculations, more recently several mixed or intermediate scenarios have been
suggested as well \cite{houdayer:00a,krzakala:00,white:06}. Still, unequivocal
(numerical or analytical) evidence for any of these scenarios could so far not be
produced.

Another, possibly even more fundamental, issue regards the question of the existence
of a spin-glass phase transition for specific spin symmetries and spatial dimensions
and, in particular, the determination of the lower critical dimension $d_l$ of the
spin-glass transition. It has been rather clear since a number of years that the
Edwards-Anderson Ising model undergoes a spin-glass transition in three dimensions
\cite{kawashima:96,ballesteros:00}, but there is no spin-glass phase at non-zero
temperatures in two dimensions \cite{bhatt:88}, such that $2\le d_l < 3$. On the
other hand, it is only recently that a consensus emerges as to the lower critical
dimension for vector spin glasses. A number of large-scale studies has consistently
shown the occurrence of a spin-glass transition at finite, but rather low
temperatures for the {\em XY\/} \cite{pixley:08} and Heisenberg spin glasses
\cite{lee:03a,campos:06,lee:07,viet:09} in three dimensions. Less agreement has been
achieved as to the question of the influence of the symmetry group O($n$), $n>1$, on
the ordering of these continuous-spin systems. Following initial suggestions by
Villain \cite{villain:79a} about the possibility of a decoupling of the Ising like
degrees of freedom embedded according to $\operatorname{O}(n) =
\operatorname{SO}(n)\oplus\mathbb{Z}_n$ in the order parameter, there is an ongoing
debate as to whether two consecutive transitions, corresponding to an ordering of
chiral and spin degrees of freedom, can be observed in these systems
\cite{weigel:05f,pixley:08,lee:03a,campos:06,lee:07,viet:09}.

In the absence of clear-cut field-theoretic and other perturbative approaches for
these systems in the relevant regime $d < d_u = 6$, much of the progress in the field
has relied on the development of new computational techniques and the use of
cutting-edge computing hardware for performing extensive computer simulation
studies. On the algorithmic side, these included the deployment of the replica
exchange or parallel tempering technique \cite{geyer:91,hukushima:96a} as well as,
for continuous-spin glasses, the adaptation of the over-relaxation move
\cite{campos:06} from its original use in lattice field theory. Besides using the
most powerful supercomputers available at each time, the hardware side of this
simulational attack to the problem has repeatedly seen the design and construction of
special-purpose machines, specifically optimized for the type of simulations at
hand. The most recent such enterprise has been the JANUS machine, a special-purpose
computer for the simulation of discrete spin systems (mostly used for spin glasses)
based on field-programmable gate arrays (FPGA) \cite{belleti:09} (see also the review
article of the JANUS collaboration in this issue). While such machines are highly
efficient, the effort for their design and construction is significant. As an
alternative route, one might hence think of using less exotic components that
nevertheless provide a massively parallel computing environment. Such hardware is
given by current-generation graphics processing units (GPUs). In a number of previous
studies, we have investigated the suitability of such devices for the general purpose
of simulating lattice spin models
\cite{weigel:10c,weigel:11,weigel:10b,weigel:10a}. One of the observations gleaned
from this experience is that within this class of applications a near ideal
application for such devices is (a) the study of disordered system due to the
inherent trivial parallelism over disorder realizations and (b) systems with
continuous spins, since these allow to harvest the large floating-point performance
and the benefits of special-function implementations in hardware. The purpose of the
present study is to see whether the 3D Heisenberg spin-glass model lives up to its
name in this respect and a GPU simulation can be performed with outstanding
efficiency.

\section{Model and method \label{sec:model}}

We consider the nearest-neighbor Edwards-Anderson spin glass model in an external,
random magnetic field,
\begin{equation}
  \label{eq:hamiltonian}
  {\cal H} = -\sum_{i,j} J_{ij}\,{\bm s}_i \cdot {\bm s}_j
  - \sum_{i} {\bm H}_i \cdot {\bm s}_i.
\end{equation}
Here, the indices $i$ and $j$ enumerate the sites of a simple (hyper-)cubic lattice
of edge length $L$ and the ${\bm s}_i$ are $m$-component classical vectors of unit
length. For the purpose of the present paper, we assume $m=3$, i.e., Heisenberg spins
and $d=3$ space dimensions, but it should be clear that our code can be adapted to
more general situations with rather modest modifications. One exception to this rule
is the heat-bath update discussed below, which cannot be formulated in closed form
for arbitrary numbers $m$ of spin components. To avoid this problem, however,
alternative techniques such as the fast linear algorithm of Ref.~\cite{loison:04} can
be adapted instead. We further assume nearest-neighbor interactions, which will be
important from the algorithmic point-of-view for the efficient use of the massive
parallelism available from GPUs using domain decompositions. The nearest-neighbor
couplings $J_{ij}$ are quenched, precomputed random variables drawn from a Gaussian
distribution with zero mean and standard deviation unity: $J_{ij}\sim {\cal N} (0,1)$
(generalizations to other nearest-neighbor $J_{ij}$ amounts to their precomputation
according to a different distribution).  Likewise, the on-site magnetic fields ${\bm
  H}_i$ are drawn independently from a uniform distribution on an $m$ dimensional
sphere of fixed radius $|{\bm H}|$. This choice is motivated by the recent work
\cite{sharma:10} suggesting that considering fields with a random {\em direction\/}
(in addition to possibly having a random magnitude) will exhibit a de
Almeida--Thouless line (at least) in the mean-field limit of such models, which is in
contrast to the findings for vector-spin systems in a uniform field that, instead,
exhibit a Gabay--Tolouse line with spin-glass ordering only in the transverse
components.

We study these systems with a combination of algorithms that is currently considered
the best tool-set for the simulation of (continuous) spin glasses (see, e.g.,
Refs.~\cite{campos:06,lee:07,viet:09}). Importance sampling is performed using
heat-bath updates, as these avoid the extra tunable parameter in form of a
spin-reorientation amplitude required for Metropolis updates of continuous spins and
they are also more efficient in terms of reduced autocorrelation times
\cite{loison:04}. The update generates a new spin state ${\bm s}'$ on site $i$
independently of the old state ${\bm s}$ according to the conditional Gibbs
distribution for this spin subject to all other spins in the system fixed:
\begin{equation}
    \label{eq:heat_bath}
    {\bm s}' \sim C e^{\beta {\bm H}_{\rm eff} \cdot {\bm s}},
\end{equation}
where $\beta$ is the inverse temperature, $C$ is a normalization constant and ${\bm
  H}_{\rm eff}=\sum_{j} J_{ij} {\bm s}_j+{\bm H}$ is the local molecular field. The
Cartesian components of ${\bm s}'$ can be explicitly obtained from two random numbers
distributed uniformly between 0 and 1, see, e.g., the discussion in
Ref.~\cite{lee:07}. It has been shown in Ref.~\cite{campos:06} that an additional
precession of the spins around their local, instantaneous molecular fields can help
to significantly speed up the decorrelation in the system. These over-relaxation
moves are microcanonical, i.e., they preserve the total energy, and can be
implemented very efficiently as they do not involve random numbers
\cite{bernaschi:10}. The most efficient implementation and, arguably, the most
efficient decorrelation of the states is achieved by simply {\em reflecting\/} each
spin at its local molecular field,
\begin{equation}
  \label{eq:microcanonical}
  {\bm s}' = 2\, \frac{{\bm s} \cdot
    {\bm H}_{\rm eff}}{{\bm H}_{\rm eff}^2} \, {\bm H}_{\rm eff} - {\bm s}.
\end{equation}
The ratio of the number of over-relaxation to heat-bath updates is an optimization
parameter to be tuned to reach the maximal decorrelation effect per clock cycle.  As
spin-glass systems suffer from notoriously slow dynamics at low temperatures, the
equilibrium of these models can hardly be studied for more than the smallest system
sizes without further, generalized-ensemble techniques for coping with their complex
free-energy landscapes. One efficient approach, that we chose to employ here, is the
parallel tempering or replica exchange method \cite{geyer:91,hukushima:96a}. There,
for every choice of $J_{ij}$ a number $R_T$ of copies of system
(\ref{eq:hamiltonian}) are simulated at a set of different, but close neighboring
temperatures $T_i$. After each copy undergoes a number of local spin updates, for
instance the combination of heat bath and over-relaxation moves discussed above, a
sweep of attempts to swap pairs $i$ and $i+1$ of neighboring configurations with
energies $E_i$ and $E_{i+1}$ at temperatures $T_i$ and $T_{i+1}$ starting from the
lowest temperature $T_1$ is made, and each swap is accepted according to the
Metropolis criterion
\begin{equation}
  \label{eq:parallel_tempering}
  p_{\rm acc}(i \leftrightarrow i+1) = \min\left[1,\,e^{\Delta \beta\Delta E}\right].
\end{equation}
Here, $\Delta E = E_{i+1}-E_{i}$ and $\Delta \beta = 1/T_{i+1}-1/T_{i}$. With an
appropriate choice of temperatures, this update allows for configurations with slow
dynamics at low temperatures to successively diffuse to high temperatures,
decorrelate there and return back to the low-temperature phase, likely arriving in a
different valley of a complex free-energy landscape. The choice of temperatures as
well as the local frequency of swap attempts can be optimized to ensure an optimal
decorrelation effect \cite{katzgraber:06,bittner:08,hasenbusch:10}.

To study spin-glass phase transitions undergone by such systems, due to the severely
restricted system sizes one is able to equilibrate, some effort needs to be invested
in devising well-suited finite-size scaling analyses. One of the most successful
approaches for spin glasses and other disordered systems \cite{ballesteros:98a}
concentrates on the role of the finite-size spin-glass correlation length
$\xi_L$. This is determined from the Edwards-Anderson order parameter,
\begin{equation}
  \label{eq:sg_order_parameter}
  q^{\mu\nu}=\frac{1}{N} \sum_i s_i^{\mu (1)} s_i^{\nu (2)},
\end{equation}
where $\mu,\nu$ are spin components and ``(1)'' and ``(2)'' denote two copies of
model (\ref{eq:hamiltonian}) with identical interactions which are simulated in
parallel. We first consider the case of a vanishing external magnetic field and
define the wave-vector dependent overlap as
\begin{equation}
  q^{\mu\nu}({\bm k}) = {1 \over N} \sum_i s_i^{\mu (1)} s_i^{\nu (2)} e^{i {\bm k} \cdot {\bm R}_i}.
\end{equation}
This leads to the ${\bm k}$-dependent spin glass susceptibility
\begin{equation}
\label{eq:susceptibility}
  \chi_{\rm SG}({\bm k}) = N \sum_{\mu,\nu} [\langle \left|q^{\mu\nu}({\bm k})\right|^2 \rangle ]_J,
\end{equation}
where $\langle \cdots \rangle$ denotes a thermal average and $[\cdots ]_J$ denotes an
average over disorder. The standard second-moment definition of the correlation
length is then given by
\begin{equation}
    \label{eq:xi}
    \xi_L = {1 \over 2 \sin (k_{\rm min}/2)} \left({\chi_{\rm SG}(0) \over \chi_{\rm SG}({\bm k}_{\rm min})} - 1\right)^{1/2}
\end{equation}
with ${\bm k}_{\rm min} = (2\pi/L)(1, 0, 0)$. To allow for a more detailed analysis
of $\chi_{\rm SG}({\bm k})$ \cite{leuzzi:09} we record a small number of additional
modes. For the case of non-vanishing magnetic fields ${\bm H}_i$, one needs to
subtract the disconnected part of the spin-glass correlation function such that
\cite{katzgraber:09}
\begin{equation}
  \chi_{\rm SG}({\bm k}) = \frac{1}{N} \sum_{i,j}\sum_{\mu,\nu} [(\chi_{ij}^{\mu\nu})^2e^{i{\bm k}\cdot ({\bm
      R_i}-{\bm R_j})}]_J,
\end{equation}
where
\[
  \chi_{ij}^{\mu\nu} = \langle s_i^\mu s_j^\nu\rangle-\langle s_i^\mu\rangle \langle s_j^\nu\rangle.  
\]
In practice, we estimate $(\chi_{ij}^{\mu\nu})^2$ by simulating in parallel {\em
  four\/} distinct real replicas with the same disorder configuration, using samples
from the time series to compute the averages
\begin{equation}
  \begin{split}
    \chi_1({\bm k}) &= \langle \sum_{i,j} \sum_{\mu,\nu} s_i^{\mu(1)} s_j^{\nu(1)} s_i^{\mu(2)} s_j^{\nu(2)}
    e^{i{\bm k}\cdot ({\bm R_i}-{\bm R_j})} \rangle, \\ 
    \chi_2({\bm k}) &= \langle \sum_{i,j} \sum_{\mu,\nu} s_i^{\mu(1)} s_j^{\nu(1)} s_i^{\mu(3)} s_j^{\nu(4)}
    e^{i{\bm k}\cdot ({\bm R_i}-{\bm R_j})} \rangle, \\ 
    \chi_3({\bm k}) &= \langle \sum_{i,j} \sum_{\mu,\nu} s_i^{\mu(1)} s_j^{\nu(2)} s_i^{\mu(3)} s_j^{\nu(4)}
    e^{i{\bm k}\cdot ({\bm R_i}-{\bm R_j})} \rangle, \\ 
  \end{split}
  \label{eq:four_replica}
\end{equation}
and assembling them at the end of the simulation to yield
\begin{equation}
  \label{eq:four_replica_two}
\chi_{\rm SG}({\bm k}) = \frac{1}{N} [\chi_1({\bm k}) -2\chi_2({\bm k})+\chi_3({\bm k})]_J.
\end{equation}
The correlation length itself is calculated from the same Eq.~\eqref{eq:xi} as for
the zero-field case.

In contrast to ferromagnets, where the symmetry group of the order parameter is
$S^n$, the $n$-dimensional unit sphere, the lack of long-range order in spin glasses
results in a symmetry group of $O(n)$, including proper as well as improper
rotations. Correlations in the embedded reflection degrees of freedom, or
chiralities, are not directly captured by the (spin) correlation length
\eqref{eq:xi}. To examine the possibility of a decoupling of spin and chiral degrees
of freedom, we therefore additionally consider the chiral correlation length which is
sensitive to such correlations. Following Ref.~\cite{olive:86}, we define the local
chirality in terms of three spins on a line as
\begin{equation}
  \kappa_i^\alpha = {\bm s}_{i+\hat{\alpha}} \cdot {\bm s}_i\times {\bm s}_{i-\hat{\alpha}}.
\end{equation}
Here, $\hat{\alpha}$ denotes one of the basis vectors of the lattice's unit cell. In
analogy to the spin degrees of freedom, we then define the chiral overlap,
\begin{equation}
  \label{eq:chiral_overlap}
  q^{\alpha}_c({\bm k}) = {1 \over N} \sum_i  \kappa_i^{\alpha (1)} \kappa_i^{\alpha (2)} e^{i {\bm k} \cdot {\bm R}_i},
\end{equation}
and the corresponding, wave-vector dependent chiral susceptibility for zero magnetic field,
\begin{equation}
  \label{eq:chiral_susceptibility}
  \chi^\alpha_{\rm CG}({\bm k}) = N [\langle \left|q_c^{\alpha}({\bm k})\right|^2 \rangle ].
\end{equation}
This, finally, allows us to define the chiral correlation length as
\begin{equation}
  \label{eq:chiral_correlation_length}
   \xi^\alpha_{c,L} = {1 \over 2 \sin (k_{\rm min}/2)} \left({\chi^\alpha_{\rm CG}(0) \over \chi^\alpha_{\rm CG}({\bm k}_{\rm min})} - 1\right)^{1/2}.
\end{equation}
To the extent it is affected by external fields, the chiral correlation length for
the case of non-vanishing fields can be estimated from four-replica expressions
analogous to Eq.~\eqref{eq:four_replica}. In addition, we consider some more
fundamental quantities such as the internal energy $U_{L}(T)=[\langle {\cal H}
\rangle ]$ and specific heat $C_{L}(T)=1/T^2[\langle {\cal H}^2 \rangle - {\langle
  {\cal H} \rangle}^2]$. Further observables, for instance the spin-glass Binder
cumulants, can also easily be measured additionally with only small computational
overhead, but we will not discuss them here.

\section{Implementation\label{sec:implementation}}

The locality of interactions, the natural prevalence of relatively expensive
floating-point operations such as those required for the heat-bath update, and the
large number of fully or nearly independent systems introduced by the required
average over disorder realizations and the parallel-tempering update, make the
present problem near ideally suited for simulations on the massively parallel
architecture provided by current GPU devices. Building on our previous experience
with simulating spin models on GPU \cite{weigel:10a,weigel:11}, we here concentrate
on a CUDA implementation for NVIDIA devices, although a more general OpenCL
realization would be feasible along rather similar lines. While parallelization over
(nearly) independent system replicas is almost trivial to achieve algorithmically,
the performance on GPU devices is, at least for the problem at hand, strongly
dominated by the efficiency of memory accesses. Hence, significant effort needs to be
devoted to the optimal memory layout of the required data structures, the
minimization of memory accesses, the adherence to the preferred memory access
patterns of the GPU devices (coalescence and avoidance of bank conflicts), and the
efficient use of the intrinsic memory hierarchy of global memory, shared memory and
registers. Technical details about these issues can be found, for instance, in
Refs.~\cite{kirk:10,cuda}. Naturally, the completely independent simulations of
different disorder realizations are easily parallelized also over different GPU
devices, which we indeed use in practice. For simplicity, however, we confine our
discussion in the present article to the description of a single-GPU code.

\subsection{Double checkerboard decomposition \label{sec:double_checkerboard}}

\begin{figure}[tb]
\centering
\includegraphics[width=0.45\textwidth]{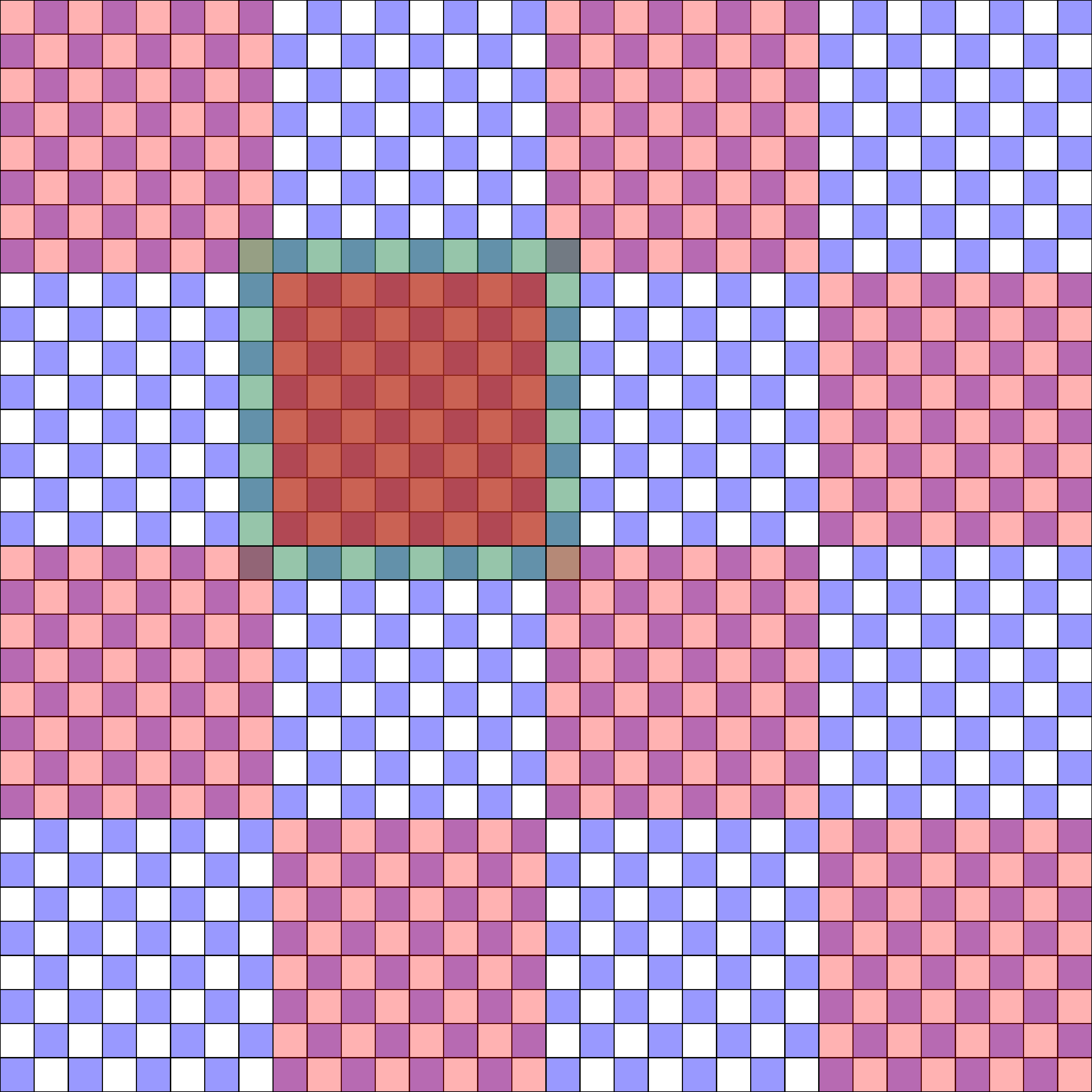} \hspace*{0.25cm}
\includegraphics[width=0.45\textwidth]{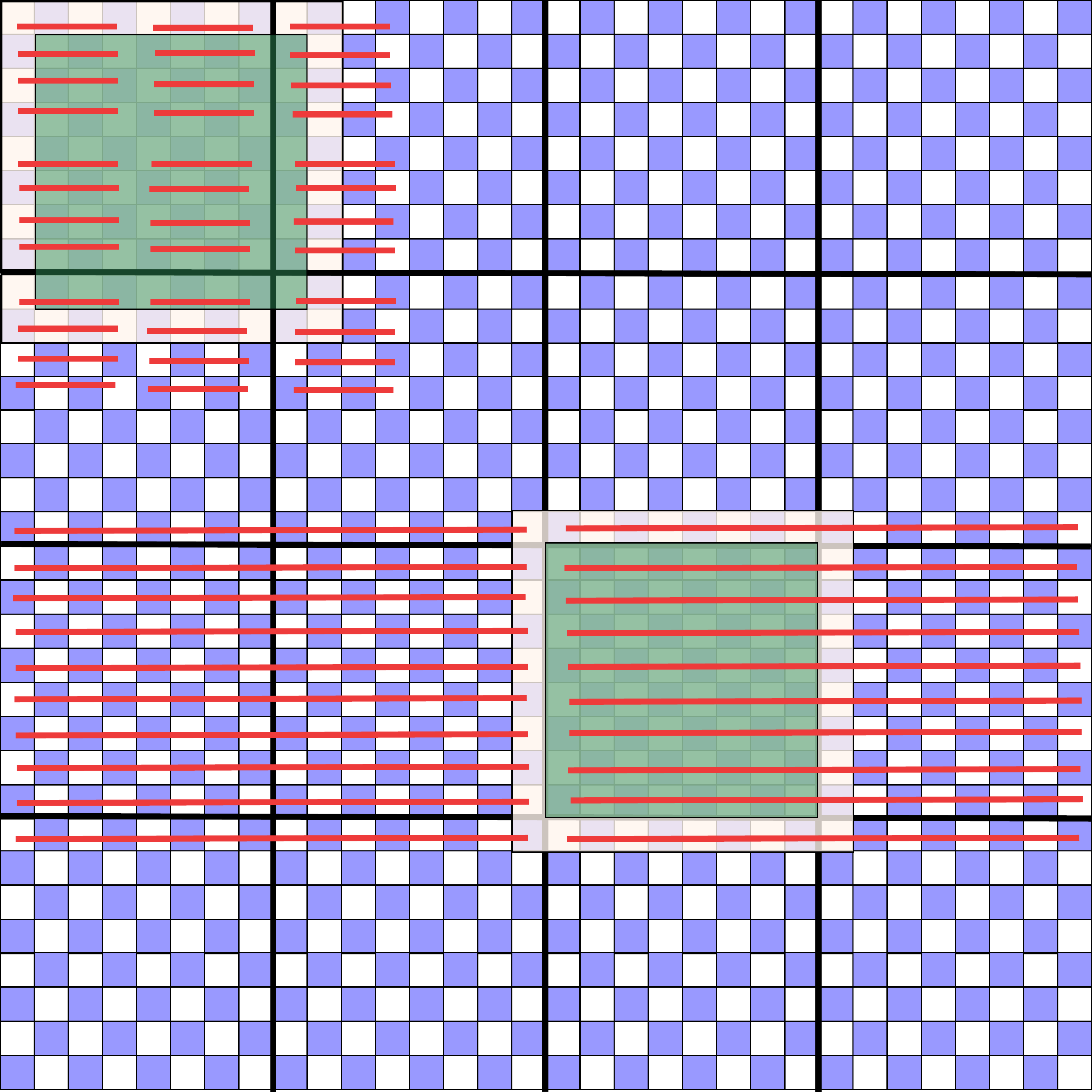}
\caption{Left panel: the double checkerboard decomposition applied to a square
  lattice of edge length $L=32$. Here, each of the $B\times B = 4\times 4$ big tiles
  is assigned as a thread block to a multiprocessor, whose individual processors work
  on one of the two sub-lattices of all $T\times T = 8\times 8$ sites of the tile in
  parallel. An analogous procedure is used for the three-dimensional spin-glass model
  considered here. Right panel: A non-linear arrangement of spins in the global
  memory address space can save on global-shared memory transactions.  The bottom
  tile covered with warp-size-long lines cartoonifies the case when memory is
  arranged linearly, and so one needs $2(8+2) = 20$ memory fetches to fetch all spins
  of the tile including halo.  In the upper left tile each long line is converted
  into a $4 \times 4$ memory tile. As a result, one needs only 9 memory fetches.  See
  text for full discussion.  }
\label{fig:checker2}
\end{figure}

The locality of interactions and spin updates allows us to use a straightforward
domain decomposition. For a bipartite lattice, such as the cubic lattice of model
(\ref{eq:hamiltonian}), the natural way of achieving this is through a (generalized)
checkerboard decomposition: all spins on one of two sub-lattices (referred to, for
instance, as ``even'' and ``odd'') can be updated in parallel without any need for
inter-thread communication. This is illustrated schematically in the left panel of
Fig.~\ref{fig:checker2}. To make efficient use of the aforementioned memory hierarchy
of GPU devices and, in particular, the availability of a small amount (currently up
to 48 kB) of shared memory with access latencies around 100 times less than those for
accessing global memory, this updating scheme is combined with a caching technique
applied to tiles which we have dubbed a {\em double checkerboard decomposition\/}
\cite{weigel:10c,weigel:10a}. There, the spin configuration of one of $B^d$ tiles
(plus a boundary layer) is loaded collaboratively into shared memory and assigned for
updating to a thread block. The threads of the block then update the even tile spins
in parallel, followed by an update of the odd spins. If these tiles are, in turn,
arranged in a checkerboard fashion (leading to the second checkerboard level),
updating of tiles of one of two sub-lattices of tiles can occur in parallel without
any communication overheads. While the use of shared memory leads to a performance
improvement as compared to any non-tiled code for a single system, the tiling turns
out to be most beneficial if {\em several\/} updates are performed on each tile
before proceeding to loading the next tile \cite{weigel:10a}. For the present model,
we perform several rounds of over-relaxation and heat-bath updates for each tile to
ensure an optimal amortization of the tile load overhead.

\subsection{General code arrangement \label{sec:general_code}}

We first give a general outline of the code and the compute kernels involved before
moving on to discussing some of the used optimizations in more detail in the
following section. From a very abstract point-of-view, our compute model works as
follows: a CPU thread calls one CUDA enabled GPU device for the concurrent simulation
of, say, $R$ copies of the system with Hamiltonian (\ref{eq:hamiltonian}). As we will
show below, this multitude of copies results from the average over disorder plus the
parallel tempering update. Each CPU thread then goes through many identical,
independent and repeatable execution rounds each consisting of three stages. First, a
CPU thread reads from the hard drive initial configurations of spins ${\bm s}$,
interactions $J$ and external magnetic fields ${\bm H}$ that describe $R$ systems, as
well as states of random number generators (RNG) and temperature points $T$ for
parallel tempering. Second, the CPU thread moves the arrays to the global memory of
the GPU device and controls execution of GPU kernels that are responsible for the
generation of new spin configurations as well as measurements of the relevant
quantities. Third, the CPU thread moves final spin configurations and states of RNG
back to the host's hard drive. The final spin configurations and RNG states are now
ready to be used as inputs for the next execution round. Moreover, breaking the
simulation into the independent runs described above permits to run, after each
round, a statistical analysis of the accrued time series and schedule for simulation
only those configurations that have not achieved convergence, as well as adjust
temperature points, if the round belongs to the equilibration stage.  Additionally,
such modularity permits a complete control over execution disruptions.

The number $R$ of copies simulated in parallel consists of $R_D$ realizations of the
disorder in the couplings and fields, times $R_O=2$ copies for each system required
to calculate the overlaps of Eqs.~\eqref{eq:sg_order_parameter},
\eqref{eq:chiral_overlap}, each of which again is multiplied by $R_T$ replicas at
different temperatures used for parallel tempering. We, therefore, represent the
system copies as a three-dimensional array of $R = R_T R_O R_D$ elements labeled by
an index\footnote{Note that we implement all multidimensional arrays flattened to
  one-dimensional C style arrays.} ${\bm r}=(r_T,r_O,r_D)$. As $R$ copies are
simulated independently, choosing a particular order of indices in ${\bm r}$ is not
crucial. However, for the parallel tempering kernel (see below) it is beneficial to
have the index $r_T$ changing slowest among the three in the energy array. The memory
organization of the data associated to an individual replica is as follows: the spin
configuration consists of $mN$ single-precision floats labeled by a four-dimensional
index ${\bm i}=(\mu,{\bm x})$, where $\mu$ spans $m=3$ spin components and ${\bm
  x}=(z,y,x)$ is the $d=3$ dimensional space index changing first, in analogy to
representing digits in a base-system number representation.  In order to benefit from
the double checkerboard decomposition, it is however convenient to think of each
system as consisting of subsystems each represented by a tile.  In this case, we
consider the index ${\bm x}$ as consisting of two sub-indices, ${\bm x}=({\bm X},
{\bm {\xi}})$, with ${\bm X}$ enumerating tiles and ${\bm \xi}$ enumerating spins
within each tile.  Hence, the spin configuration associated to a replica is organized
in a $10=3+7$-dimensional array with index $i\equiv ({\bm r},{\bm i})\equiv({\bm
  r},\mu,{\bm X},{\bm \xi})$. Similarly, the couplings and random fields require each
8-dimensional indices.

The updating of spins is performed entirely on GPU. A first kernel, doing the {\em
  local updating\/} performs the heat-bath and over-relaxation moves. It utilizes the
double-checkerboard decomposition discussed above. If a checkerboard with $B^d$
coarse tiles is chosen, we launch $B^d/2$ thread blocks per replica working on all
the tiles of one sub-lattice in parallel. On each tile, we perform $n_M$
microcanonical over-relaxation sweeps and $n_B$ heat bath sweeps; we typically take
$n_M=10$ and $n_B=1$ \cite{lee:07}. Looping over sub-lattices, we update the whole
lattice $n_T$ times, after which we attempt a parallel tempering move, with,
typically, $n_T=1$. A second kernel performs the {\em parallel tempering update\/}. A
thread in a block of this kernel performs all necessary temperature exchanges for a
given disorder realization $r_D$ and a given overlap index $r_O$. Such calculations
are independent and $R_D R_O$ threads necessary for all temperature exchanges can be
distributed among blocks rather arbitrarily. The energies required for
Eq.~\eqref{eq:parallel_tempering} are reconstructed in the parallel-tempering kernel
from energy increments recorded continuously in the course of the heat-bath
updates. After the replica exchange has been performed $n_A$ times, the quantities of
interest, such as the current energy and the susceptibilities $\chi_{\rm SG}({\bm
  k})$ and $\chi^\alpha_{\rm CG}({\bm k})$ are calculated.  For instance, to obtain
the ${\bm k}$-dependent spin glass susceptibility, Eq.~(\ref{eq:susceptibility}),
entries of the time series of $\chi_1$, $\chi_2$, $\chi_3$ of
Eq.~\eqref{eq:four_replica} are generated for the $R$ copies simulated concurrently
and immediately copied to host memory. Since measurements are performed relatively
rarely, the overhead of such memory transfers is not an issue here. One complete
round of the Monte Carlo simulation is finished as soon as a time series of length
$T_{\rm prod}$ for each physical quantity has been accrued. In total, the Monte Carlo
procedure on GPU thus looks as follows:
\begin{enumerate}
\item \label{alg:1} The {\it local updating} kernel is launched with $B^2/2 \times
  \NS$ thread blocks assigned to treat the {\it even} tiles of the coarse
  checkerboard in each of $\NS$ systems.
\item \label{alg:2}
The $T^2/2$ threads of each thread block cooperatively load into shared memory the 
configuration of spins ${\bm s}$, couplings $J$ and external magnetic fields ${\bm H}$ of their tile,
plus a boundary layer (referred to as the ``halo'' below).
\item \label{alg:3}
The threads of each block perform an over-relaxation update of each {\it even}
lattice site in their tile in parallel.
\item \label{alg:4}
All threads of a block wait for the others to finish at a barrier synchronization point.
\item \label{alg:5}
The threads of each block perform an over-relaxation update of each {\it odd}
lattice site in their tile in parallel.
\item \label{alg:6}
The threads of each block are again synchronized.
\item \label{alg:61}
Steps \ref{alg:2}--\ref{alg:6} are repeated $n_M$ times.
\item \label{alg:7}
Steps \ref{alg:2}--\ref{alg:6} are repeated $n_B$ times, with
over-relaxation updates replaced by heat-bath updates.
\item \label{alg:8}
A second {\it local updating} kernel is launched, repeating steps
\ref{alg:1}--\ref{alg:7} on the {\it odd} tiles of the coarse checkerboard 
of $\NS$ systems.
\item \label{alg:9}
Steps \ref{alg:1}--\ref{alg:8} are repeated $n_T$ times.
\item \label{alg:10}
A {\it parallel tempering} kernel is executed permuting temperatures on $\NS$ systems.
\item \label{alg:11}
Steps \ref{alg:1}--\ref{alg:10} are repeated $n_A$ times.
\item \label{alg:12}
{\it Analysis kernels} are launched to calculate physical quantities of interest.
\item \label{alg:13} Steps \ref{alg:1}--\ref{alg:12} are repeated $T_{\rm prod}$
  times, and can be preceded by $T_{\rm equi}$ steps for equilibration without
  launching {\it analysis kernels} in step \ref{alg:12}.
\end{enumerate}
We note that $n_M$, $n_B$, $n_T$, $n_A$ and $R_T$ can be considered as optimization
parameters, minimizing autocorrelation times of physical quantities of interest in
units of $t_{\rm update}$, while $T_{\rm prod}$ can be chosen to make $t_{\rm run}$
of the order of hours, see Table \ref{tab:speed-up} and the accompanying discussion
for a definition of $t_{\rm run}$ and $t_{\rm update}$.

Random numbers are required for performing the heat-bath update. For the purpose of
the benchmark runs reported below, we used the multiply-with-carry generator first
implemented on GPU in Ref.~\cite{alerstam:08}. While the statistical quality of the
thus generated sequences is not outstanding, it is presumably sufficient for the
purpose at hand, in particular since the quenched couplings and fields are generated
off-GPU with the help of another, high-quality RNG. Since the proportion of the total
run-time consumed by random-number production is small compared to the remaining
arithmetic, in particular since the over-relaxation moves employed generously do not
consume random numbers, we are planning to replace the multiply-with-carry algorithm
with another generator of higher quality, see also the review on random-number
generators on GPU in this issue \cite{manssen:12}.

\subsection{Memory optimizations}

In the described algorithm $R_T R_O$ out of $\NS$ systems are simulated with the same
realizations of couplings $J$ and external magnetic fields ${\bm H}$. To avoid
unnecessary transfers between global and shared memory, we assign to each block a
number $R_{\overline{TO}}\geq 1$ of different systems that are sequentially simulated
with the same configuration of $J$ and ${\bm H}$.  In this case, in the algorithmic
steps \ref{alg:1} and \ref{alg:8} above, only $(B^2/2 \times \NS)/R_{\overline{TO}}$
blocks are scheduled for execution, while steps \ref{alg:2}--\ref{alg:7} are
sequentially repeated on $R_{\overline{TO}}$ systems with different ${\bm s}$ but the
same $J$ and ${\bm H}$, allowing for loading the latter only once.  We note that
$R_{\overline{TO}}$ can be deemed yet another optimization parameter.

The highest bandwidth of memory transactions between global and shared memory is
achieved for coalesced memory fetches \cite{kirk:10}, i.e., accesses conforming to
certain pre-described patterns. More precisely, global memory loads and stores by
threads of a warp can be coalesced by the device into one memory transaction,
provided that all 32 threads of a warp access 4-byte words (floats) located in the
same 128-bit aligned memory segment\footnote{This is for devices of compute
  capability 2.x which we have used here.}. To organize cooperative loading of each
tile along with its halo into shared memory in a fashion to ensure coalescence
wherever possible, a non-linear storage of arrays for ${\bm s}$, $J$ and ${\bm H}$ in
the address space of global memory might be beneficial. Specifically, a much higher
ratio of coalesced to non-coalesced memory transactions can be achieved by suitably
positioning tiles on the lattice and by adding additional granularity to the
spin-enumerating index ${\bm r}$, in each out of $\NS$ lattices. This is illustrated
in the right panel of Fig.~\ref{fig:checker2} for a square lattice. For simplicity,
we assume here that the state of each spin only occupies a single 32-bit word. Also,
for the sake of argument, assume that the warp size is 16 and the tile size is $16
\times 16$.  If the floats for $s_i$ are stored sequentially in global memory and
tiles are aligned to the edges of the lattice, fetching every tile {\it plus} its
halo amounts to 20 memory transactions. This is the case for the lower tile in the
right panel of Fig.~\ref{fig:checker2}, where linearly aligned warp-length addresses
are shown with long red lines.  Note that fetching just one float for the left halo
is as time consuming as fetching all 16 floats would have been.  Contrary to that, if
one stores floats in a non-linear fashion, for example, in the form of $4\times 4$
{\it memory tiles} and if one misaligns the lattice tile, fetching every such tile
{\it plus} its halo amounts to merely 9 memory transactions as shown for the upper
left tile. There, floats are aligned in a sequential manner first within small {\it
  memory tiles}, such that each square consisting of four red lines of length four
represents sequential addresses, and then the memory tiles are themselves positioned
sequentially.  In addition, the lattice tile is shifted by $1$ both in $x$ and $y$
directions.  Note, that floats in every memory transaction, represented by a memory
tile, are now used more economically. The same type of argument can be made for the
three-dimensional lattices and $m$ floats per spin for the model actually under
consideration.

After loading the tile into shared memory we intend to perform several sweeps over
the whole tile, see the algorithm above. Updating each spin requires calculation of
the local magnetic field vector ${\bm H}_{\rm eff}=\sum_{j} J_{ij} {\bm s}_j+{\bm
  H}$, and thus accessing neighboring interactions and states of neighboring spins.
This can be done most efficiently by using the shared memory address space in a {\it
  linear} fashion, unlike in the global memory.  This means that while introducing
global memory tiles saves on memory transactions to shared memory, it entails an
overhead for additional index operations to restore a non-tiled layout in shared
memory. To capture the interplay between these two opposite effects we consider both
lattice tile sizes and memory tile sizes as optimization parameters.  We avoid shared
bank conflicts by padding, which reduces the percentage of conflicts down to a few
percent of the accesses.

Another optimization saving on memory bandwidth results from packing the state $(s_z,
s_y, s_x)$ of each spin into two numbers. An obvious choice would be exploiting ${\bm
  s}=1$ to use polar angles $\theta$, $\phi$.  Here, instead, we decided to store for
each spin the two numbers $s_y$ and $s_x+\Delta\,\operatorname{sign}(s_z)$, where we
chose the arbitrary offset $\Delta=4$. Restoration of the three Euclidean spin
components is then performed in shared memory with the help of only one (costly)
transcendental function.

We have also attempted an alternative arrangement of the code that does not rely on
the double-checkerboard decomposition with its heavy use of shared memory, but on a
direct fetch of data for ${\bm s}$, $J$, ${\bm H}$ from global memory to registers
(see also Ref.~\cite{bernaschi:10}). In this setup, the spin index $i$ naturally
decomposes as $i\equiv(\mu,\sigma,{\bm X},{\bm r})$, where $\sigma$ and ${\bm X}$
enumerate the two sub-lattices and $2\times 2$ tile ``super-cells'' of the cubic
lattice, correspondingly, and ${\bm r}$ changes first. The need to simulate a large
number $R$ of systems ensures coalescence of global-register fetches in such a
``vertical'' arrangement of data in global memory. This results in somewhat simpler
code, but also comes with a number of drawbacks, namely (i) the need to fetch 30
floats for every spin update (the spin, 6 neighboring spins, 6 couplings, 1 magnetic
field), or 22 floats in the packed implementation, (ii) the impossibility to recycle
$J$ and ${\bm H}$, and (iii) that $R_D$ families of systems with different disorder
realization cannot be easily considered separately (which might be necessary if one
wants to run certain realizations longer than others). We therefore abandoned this
type of code in favor of the code layout discussed above.

Apart from the {\it local updating kernel}, which is responsible for the bulk of the
GPU time, we set up a {\it parallel tempering kernel}, where each of a total of $R_D
R_O$ threads is assigned to permute temperatures in all $R_T$ systems with the same
$J$. To ensure coalescence, the matrices keeping track of permutation indices, as
well as temperature and energy arrays are stored in global memory with the
temperature index changing slowest. The division of the $R_O R_D$ threads into
thread blocks is arbitrary as they are independent, and is dictated solely by time
efficiency. In the {\it analysis kernel} for the calculation of $\chi_{\rm SG}({\bm
  k})$ and $\chi^\alpha_{\rm CG}({\bm k})$, every block of threads works on $R_O$
complete systems, by sequentially loading into shared memory and processing tiles of
spins (of not necessarily the same size as the ones used in the {\it local updating
  kernel}) to cover their entire lattices, while different blocks are scheduled $R_T
R_D$ times.  The arithmetic load of each thread block to process each tile for
several values of ${\bm k}$ and for all combinations of the spin components $\mu$,
$\nu$, see Eq.~\eqref{eq:susceptibility}, is high (even disregarding the calculation
of the sines and cosines involved), making this kernel compute bound and thus
unlikely to benefit from other mappings onto the grid-and-block execution model of
the GPU. Since measurements occur relatively infrequently, however, the impact of
this kernel on the total execution time is small.

\section{Benchmarks and testing}

It remains to confirm that the chosen implementation, in particular regarding the
mixed-precision computations with spins being represented in single precision and
aggregate quantities such as measured correlation functions in double precision, is
correct and provides results in agreement with reference CPU implementations. To this
end, we compared our GPU runs to our own CPU implementation and found statistical
consistency of the results of our final implementations. 

\begin{figure}[tb]
  \centering
  \includegraphics[width=0.8\textwidth]{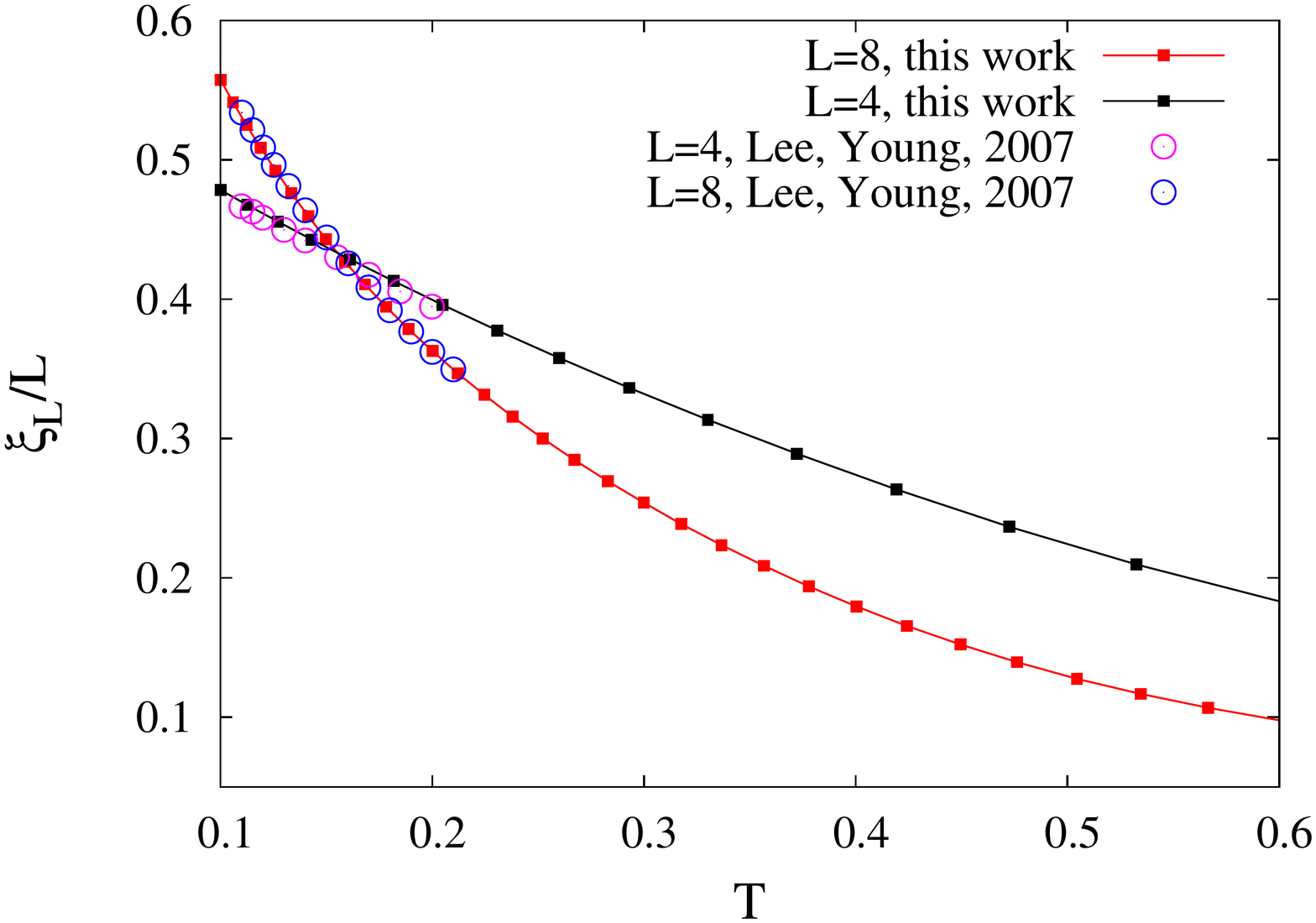} 
  \caption{ The spin-glass correlation length $\xi_L$, Eq.~(\ref{eq:xi}), divided by
    system size $L$ as a function of temperature for $L=4$ and $8$. Our results
    reproduce those of Ref.~\cite{lee:07} well. The simulation parameters were chosen
    as $R_T=32$ ($L=8$) and $R_T=16$ ($L=4$), $R_O=2$, $T_1=0.1$, $T_{R_T}=0.6$,
    $n_A=10$, $n_T=1$, $n_M=10$, and $n_B=1$. For each disorder realization, a time
    series of length $T_{\rm prod}=2\times 10^4$ was collected, after discarding the
    initial $T_{\rm equi}=8\times 10^4$ records to allow for equilibration. $R_D=256$
    disorder realizations were studied.  }
  \label{fig:ksi}
\end{figure}

To completely convince ourselves of the correct implementation of our codes, we also
compared parts of the final results to some of the published data on the Heisenberg
spin glass. In particular, Fig.~\ref{fig:ksi} shows our estimates for the finite-size
spin-glass correlation length for system sizes $L=4$ and $L=8$ as a function of
temperature, compared to the corresponding data gleaned from Lee and Young
\cite{lee:07} for the case of zero random field. This comparison serves as an
important check as to the sufficiency of the mixed-precision ansatz for the intended
simulations.

To assess the quality of the implementation, we performed a series of benchmark runs,
comparing the GPU implementation against our CPU based reference. The results are
summarized in Table \ref{tab:speed-up}, featuring average times per spin update for
an $L=32$ system, which is realistic for the (larger of the) system sizes to be
considered in production runs. The speed-ups for larger systems are similar to and
the speedups for smaller systems somewhat smaller than those reported in Table
\ref{tab:speed-up}.  The benchmark runs were performed on the GTX 480 Fermi card,
equipped with 15 streaming multiprocessors of 32 cores each at a processor clock of
1.4 GHz; as a reference CPU, an Intel Core 2 Quad Q9650 at 3.0 GHz was used. Against
the serial code, overall speed-ups of up to a remarkable factor of 150 are
observed. Here, if the whole simulation of $\NS$ systems took physical time $t_{\rm
  run}$ in seconds on the device, Table \ref{tab:speed-up} quotes $t_{\rm
  update}=t_{\rm run}/[(n_M+n_B)n_T\,\NS\,N]$ as the time of one spin update. As a
typical mixing rate \cite{lee:07}, we considered the case of one heat-bath and ten
over-relaxation moves per sweep, resulting in an average time of 0.45 ns per update
--- which is brought into perspective comparing to the 0.50 ns for a pure Metropolis
update of a 2D Heisenberg ferromagnet reported in Ref.~\cite{weigel:10a}.

As mentioned above, one additional optimization is possible on using the fast
intrinsic special-function implementations provided on GPU. While this has no effect
on the over-relaxation moves not involving special-function calls, cf.\
Eq.~\eqref{eq:microcanonical}, the heat-bath update significantly profits from
activating this feature, cf.\ the data for ``float, fast\_math'' in Table
\ref{tab:speed-up}. Importantly, we did not observe any systematic deviations of the
final results on activating the fast, but reduced precision intrinsics. The
differences coming from packing the state $(s_z,s_y, s_x)$ into two numbers are found
to be insignificant.

In terms of parameter optimization, using the double checkerboard decomposition and
the multiply-with-carry random number generator \cite{alerstam:08}, close-to-optimal
running times of the {\it local updating kernel} are obtained for shared memory tile
sizes $8\times8\times8$ with padding, memory tile sizes $2\times2\times2$, see
Fig.~\ref{fig:checker2}, and recycling of $J$ and ${\bm H}$ at
$R_{\overline{TO}}\geq8$, see sec.~\ref{sec:implementation}. For the chosen
implementation, there is obviously a rather vast space of optimization parameters
which would be very time consuming to scan systematically. Even more, for a realistic
and fair comparison, one would require to compare all options in terms of times to
generate uncorrelated events. These can differ between the considered variants, in
particular for situations with multi-hit updates on tiles, which nominally yield
larger speed-ups (cf.\ the last two lines in Table \ref{tab:speed-up}), but will also
result in somewhat increased autocorrelation times \cite{weigel:10a}. A more complete
scan of the space of optimization parameters will be discussed elsewhere.

\begin{table}
\centering
\caption{
  Times per spin update in ns for different variants of our GPU implementation as
  compared to the CPU reference code. ``HB'' is a heat-bath update, ``OR'' refers to
  an over-relaxation move. The variants with ``fast\_math'' refer to the case of
  using the fast intrinsic special-function implementations.
  The data were generated running $R=256$ systems of size  $L = 32$. For these benchmark
  runs, we chose temperature $T=100$ to avoid any dependence on the disorder
  realization, but we checked that the results at more realistic temperatures are
  virtually identical. Note that all times are per single spin update, irrespective
  of whether it is a heat-bath or over-relaxation move.
  On CPU, execution times behave linearly; for instance,
  1 HB + 10 OR take $61\approx(242+10 \times 43)/11$ ns.
}
\begin{tabular}{l*{6}{l}} 
\hline
\multicolumn{1}{c}{device} & \multicolumn{1}{c}{mode} & 
\multicolumn{1}{c}{update} &
\multicolumn{1}{c}{$t_\mathrm{update}$ [ns]} & \multicolumn{1}{c}{speedup} 
\\ \hline
			& double		&  1 HB			& 242			& 1		\\ 
Intel Q9650	        & double		&  1 OR		& 43				& 1		\\ 
			& double		&  1 HB	+ 10 OR	& 61                      	& 1		\\ \hline
	     		& float                 &  1 HB			& 2.23                          & 108	\\
	     		& float, fast\_math   	&  1 HB			& 1.58                          & 153 	\\
     		        & float               	&  1 OR		& 1.30                          & 33	\\
     	   	        & float, fast\_math    	&  1 OR		& 1.30			        & 33    \\
GTX 480			& float                	&  1 HB + 10 OR	& 0.45		                & 136	\\
			& float, fast\_math    	&  1 HB + 10 OR	& 0.39		                & 156	\\
			& float                	&  1 HB + 100 OR& 0.29		                & 155	\\
			& float 	        &  10 $\times$ (1 HB + 10 OR)	&  0.36	        & 169	\\
			& float, fast\_math	&  10 $\times$ (1 HB + 10 OR)	&  0.31	        & 197	\\ \hline
\end{tabular}
\label{tab:speed-up}
\end{table}

\begin{figure}[tb]
  \centering
  \includegraphics[width=0.8\textwidth]{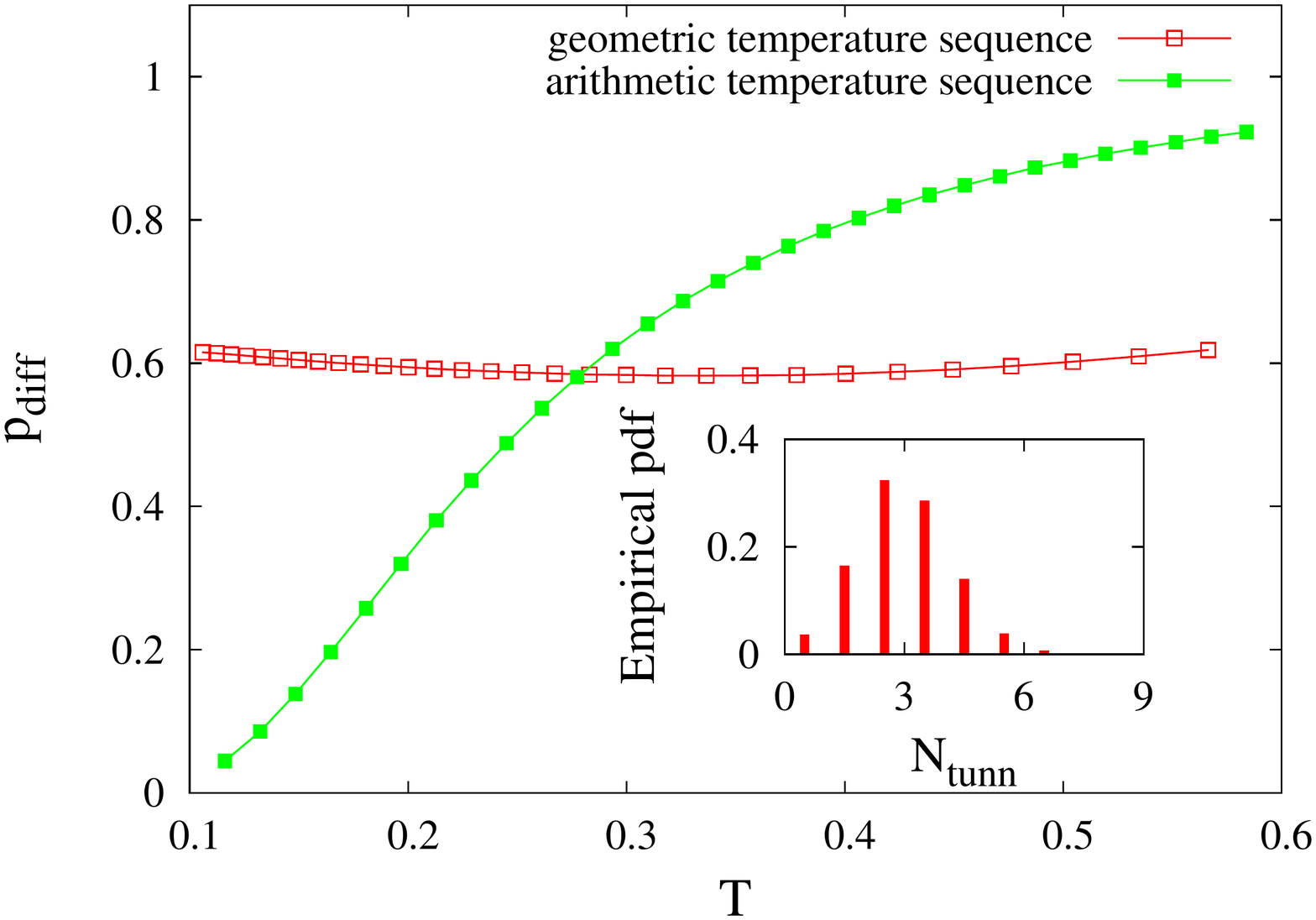} 
  \caption{ Probability $p_{\rm diff}$, averaged over disorder realizations, that a
    spin configuration at a given temperature $T$ diffuses to neighboring $T$ with
    the parallel tempering exchange algorithm for an arithmetic and a geometric
    sequence of temperature points $T_i$.  Inset: distribution of the number of
    tunneling events over $\NS = 2^{14} = 16\,384$ systems studied.  A tunneling
    event occurs when a replica travels a closed loop between the temperature end
    points $T_{1}$ and $T_{R_T}$.  The presented data are for system size $L=8$ and
    ${\bm H=0}$ in Eq.~(\ref{eq:hamiltonian}). Simulation parameters are as in
    Fig.~\ref{fig:ksi}, but with $n_A=1$.
 } 
  \label{fig:T2C-C2T}
\end{figure}

One direction in the multidimensional optimization space deserves special attention,
as it is related to the algorithmic aspect of simulations. This is a suitable choice
of the number, $R_T$ and position, $T_i$, of temperature points in a fixed
temperature interval used for the parallel tempering moves
\cite{katzgraber:06,bittner:08,hasenbusch:10}. We checked earlier observations
\cite{lee:07} that choosing $T_i$ in a geometric manner: $T_i=T_1 f^{i-1},
f=(T_{R_T}/T_1)^{1/(R_T-1)}$, in contrast to a linear spacing of $T_i$, is crucial
for allowing configurations with slow dynamics at low temperatures to successively
diffuse to high temperatures, and decorrelate.  In Fig.~\ref{fig:T2C-C2T} we show the
probability $p_{\rm diff}$ that, after a parallel tempering update, a copy of the
system at $T_i$ will diffuse to neighboring temperatures.  It is clear that the
linear choice of $T_i$ is inefficient in that it disfavors the most important
exchanges at low $T_i$. On the contrary, a geometric choice of $T_i$ results in
simulated copies of the system successfully undergoing a random walk in the whole
temperature interval, traveling a closed loop between $T_{1}$ and $T_{R_T}$ for
$N_{\rm tunn}\approx 3$ times on average during the production run of length $2\times
10^4$ sweeps, see the inset in Fig.~\ref{fig:T2C-C2T}. For the linear temperature
protocol, essentially no tunneling events are observed.

\section{Conclusions}

Spin-glass simulations are extremely CPU hungry applications. In the present paper,
we have shown that graphics processing units (GPUs) are an interesting alternative to
conventional computer clusters and special-purpose computers for the specific problem
of simulating short-range spin glasses with continuous spins. Carefully crafting the
code to reduce memory bandwidth consumption, bank conflicts and communication
overheads, and to increase concurrency, coalescence, and device load it is possible
to achieve speed-ups against serial CPU code of around 150 for this problem. While
these results are encouraging, it is still worthwhile to point out that some of the
details of the implementation are rather intricate. One example of this is our
observation of systematic drifts in energy in the energy-conserving microcanonical
moves if the fusion of multiplications and additions by the compiler is not
explicitly prevented.

One of the advantages of the chosen code layout is that the total run-times for
individual disorder realizations can be chosen independent of each other, thus
allowing to accommodate the fat-tailed distribution in ``hardness'' of disorder
realizations to be expected from spin-glass systems
\cite{moreno:03,bittner:06,weigel:06b}. This flexibility will be essential for
performing more extensive simulations of the problem considered here, which will also
put to test the reliability of the new compute model in a large-scale application.

\section*{Acknowledgments}

The authors thank E.\ Sch\"omer for his contributions at an early stage of the
project, as well as H.\ Katzgraber for enlightening discussions. The authors
acknowledge support by the ``Center for Computational Sciences in Mainz'' (SRFN),
computer time provided by NIC J\"ulich under grant No.\ hmz18 and funding by the DFG
under contract No.\ WE4425/1-1 (Emmy Noether Programme).


\end{document}